\documentclass[conference, final]{IEEEtran}

\usepackage{cite}
\usepackage[caption=false,font=footnotesize]{subfig}
\usepackage{graphicx}
\usepackage[cmex10]{amsmath}
\usepackage{amsfonts,amssymb,amsthm}
\usepackage{mathrsfs}
\usepackage{enumerate}
\usepackage{color}

\usepackage{tikz}
\usepackage{verbatim}
\usetikzlibrary{shapes.gates.logic.US,trees,positioning,arrows}

\begin{document}

\title{Performance Analysis of Random Linear Network Coding in Two-Source Single-Relay Networks}


\author{\IEEEauthorblockN{Amjad~Saeed~Khan and Ioannis~Chatzigeorgiou}
\IEEEauthorblockA{School of Computing and Communications, Lancaster University, United Kingdom\\
Email: \{a.khan9, i.chatzigeorgiou\}@lancaster.ac.uk}}

\maketitle

\begin{abstract}
This paper considers the multiple-access relay channel in a setting where two source nodes transmit packets to a destination node, both directly and via a relay node, over packet erasure channels. Intra-session network coding is used at the source nodes and inter-session network coding is employed at the relay node to combine the recovered source packets of both source nodes. In this work, we investigate the performance of the network-coded system in terms of the probability that the destination node will successfully recover the source packets of the two source nodes. We build our analysis on fundamental probability expressions for random matrices over finite fields and we derive upper bounds on the system performance for the case of systematic and non-systematic network coding. Simulation results show that the upper bounds are very tight and accurately predict the decoding probability at the destination node. Our analysis also exposes the clear benefits of systematic network coding at the source nodes compared to non-systematic transmission.
\end{abstract}

\vspace{1mm}

\section{Introduction}
\label{sec:intro}

In their seminal paper \cite{Ahlswede2000}, Ahlswede \textit{et al.} proposed network coding as a means to share resources by allowing intermediate network nodes to combine packets prior to forwarding them to the next-hop nodes. Network coding, which can improve throughput in both wired and wireless networks, is classified into \textit{intra-session} and \textit{inter-session} network coding \cite{Seferoglu2011}. The former approach mixes packets within the same flow, for example, packets of the same source node, and can be implemented using fountain codes. The latter scheme mixes packets of different flows, for example, packets that originate from different source nodes.

Network coding can be combined with wireless node cooperation to reap the benefits of spatial diversity. A system configuration that is often considered in conjunction with network coding is the multiple-access relay channel (MARC). In the case of two source nodes, source packets are transmitted to both a relay node and a destination node on orthogonal or non-orthogonal channels. The relay node linearly combines the received packets of the two source nodes using \textit{inter-session} network coding and forwards the network-coded packets to the destination node. The diversity gain and the performance of the system, measured in terms of the probability that the destination node will fail to recover the source packets of any of the source nodes, was studied by Bao and Li \cite{Bao2005} and Chen \textit{et al.} \cite{Chen2006}. Woldegebreal and Karl \cite{Woldegebreal2007} considered a similar setup to that in \cite{Bao2005,Chen2006}, and investigated the impact of two different relay operating modes on the coverage area of the system. In an effort to improve the coding gain of the MARC, joint channel coding and network coding was introduced, for example in \cite{Hausl2006,Yang2007}, while compute-and-forward was proposed in \cite{Nokleby2012} as a means to increase the achievable rate of the MARC.

An aspect that \cite{Bao2005,Chen2006,Woldegebreal2007,Hausl2006,Yang2007,Nokleby2012} have in common is that network coding is not used at the source nodes. Nevertheless, systems that employ \textit{intra-session} network coding both at the source and the relay stages have been analyzed in the literature, for example \cite{Molisch2007,Tarable2009,Kurniawan2010}. However, the aforementioned work is concerned with the transmission of \mbox{network-coded} packets from a single source to a single destination via multiple relays. Consequently, the destination is not required to recover and identify source packets of different source nodes.

This paper is concerned with transmission over an orthogonal MARC when two source nodes are present. The motivation of this work is to study a network configuration that encompasses both intra-session network coding at the source nodes, as in \cite{Molisch2007,Tarable2009,Kurniawan2010}, and inter-session network coding at the relay node, as in \cite{Bao2005,Chen2006,Woldegebreal2007}. The probability that the destination node will successfully recover the source packets of both source nodes is used as the performance measure of the system. Our analysis builds on and extends recent work on random binary block angular matrices \cite{Ferreira2013}. In our study, we have looked at the \textit{decode-and-forward} relaying scheme, that is, network-coded packets received by the relay node are decoded and re-encoded before they are forwarded to the destination node. The derived probability expressions could be adapted to other network-coded relaying strategies that incorporate both intra-session and inter-session network coding schemes, as in \cite{Fitzek2013}, or be used as benchmarks in performance comparisons.
 
The rest of the paper is organized as follows. Section~\ref{sec:model} describes the two-source single-relay system model in detail and formulates the problem statement. Section~\ref{sec:analysis} builds on fundamental probability expressions for decoding \mbox{network-coded} packets and obtains tight closed-form bounds on the probability of the destination node recovering the source packets of both source nodes. Section~\ref{sec:results} validates the theoretical analysis and discusses simulation results. The contributions of the paper and future research directions are summarised in Section~\ref{sec:concl}.

\section{System Model and Problem Statement}
\label{sec:model}

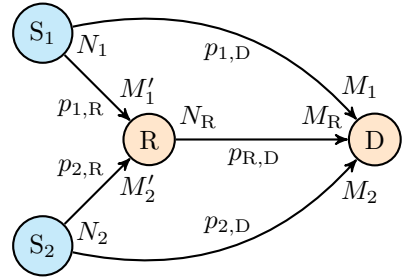
\begin{figure}[t]
\centering
\begin{tikzpicture}[->, thick, auto, on grid, >=stealth',
		source/.style={circle, draw, fill=cyan!20},
		sink/.style={circle, draw, fill=orange!20}]
	\node [sink] (d) {$\mathrm{D}$};
	\node [sink] (r) [left=3cm of d] {$\mathrm{R}$};
	\node [source] (s2) [below left=2cm of r] {$\mathrm{S}_2$};
	\node [source] (s1) [above left=2cm of r] {$\mathrm{S}_1$};
	\node at (-1.95,1.15) {$p_\mathrm{1,D}$};
	\node at (-1.95,-1.15) {$p_\mathrm{2,D}$};	
	\node at (-3.9,0.4) {$p_\mathrm{1,R}$};	
	\node at (-3.9,-0.4) {$p_\mathrm{2,R}$};	
	\node at (-1.6,-0.25) {$p_\mathrm{R,D}$};
	\node at (-3.75,1.25) {$N_1$};	
	\node at (-3.15,0.65) {$M'_1$};		
	\node at (-3.75,-1.25) {$N_2$};	
	\node at (-3.15,-0.65) {$M'_2$};	
	\node at (-2.35,0.3) {$N_\mathrm{R}$};
	\node at (-0.7,0.3) {$M_\mathrm{R}$};		
	\node at (-0.2,0.7) {$M_\mathrm{1}$};	
	\node at (-0.2,-0.7) {$M_\mathrm{2}$};						
	\path[every node/.style={font=\sffamily\small}]
    (r)  edge node {} (d)
    (s2) edge node {} (r)
         edge [bend right] node {} (d)
    (s1) edge node {} (r)
         edge [bend left] node {} (d);
\end{tikzpicture}
\caption{Block diagram of a network consisting of two source nodes $\mathrm{S}_1$ and $\mathrm{S}_2$, a relay node $\mathrm{R}$ and a destination node $\mathrm{D}$. The packet erasure probability of each link as well as the number of transmitted and received coded packets at each node are also depicted.}
\vspace{-2mm}
\label{fig:block_diagram}
\end{figure}

We consider a network comprising two source nodes $\mathrm{S}_1$ and $\mathrm{S}_2$, a relay node $\mathrm{R}$ and a destination node $\mathrm{D}$, as shown in Fig.~\ref{fig:block_diagram}. Nodes $\mathrm{S}_1$ and $\mathrm{S}_2$ segment data into $K_1$ and $K_2$ equally-sized packets, respectively. Let $u_1,\dots,u_{K_1}$ denote the source packets of node $\mathrm{S}_1$ while $u_{{K_1}+1},\dots,u_{{K_1}+{K_2}}$ represent the source packets of node $\mathrm{S}_2$. Each source node employs random linear network coding to combine  source packets and generate coded packets. In \textit{non-systematic} network coding, each source transmits ${N_\ell}\geq {K_\ell}$ coded packets, where $\ell=1,2$. In \textit{systematic} network coding, the first $K_\ell$ transmitted packets are identical to the source packets, while the remaining ${N_\ell}-K_\ell$ packets are coded. As is customary in network coding, each coded packet is transmitted along with a coding vector, which contains the $K_\ell$ coefficients of the respective linear combination. In this paper, we consider coefficients that are chosen uniformly at random from the elements of the Galois field $\mathrm{GF}(2)$. Therefore, each coded packet is the bitwise sum of source packets.

Links between network nodes are modelled as packet erasure channels. We use $p_\mathrm{\ell,D}$, $p_\mathrm{\ell,R}$ and $p_\mathrm{R,D}$ to denote the packet erasure probabilities of the links connecting the $\ell$-th source node with the destination node, the $\ell$-th source node with the relay node and the relay node with the destination node, respectively. We assume that source nodes transmit on orthogonal channels enabling both the relay and the destination nodes to distinguish transmissions between the source nodes.

The communication process is split into two phases. In the first phase, nodes $\mathrm{S}_1$ and $\mathrm{S}_2$ transmit $N_1$ and $N_2$ coded packets, respectively, to node $\mathrm{D}$. Node $\mathrm{R}$ overhears the transmissions of the source nodes, stores the successfully received coded packets and attempts to decode them. Let $M_\ell$ and $M'_\ell$ be the number of coded packets from node $\mathrm{S}_\ell$ that were received by the destination node $\mathrm{D}$ and the relay node $\mathrm{R}$, respectively. The coding vectors of the received coded packets can be stacked together at the receiving nodes to form coding matrices. At the end of the first phase, the coding matrices at nodes $\mathrm{D}$ and $\mathrm{R}$ can be expressed in block diagonal form as follows 
\begin{equation}
\label{eq:codmtx_SD_SR}
\mathbf{C}_\mathrm{SD} = \left(\begin{IEEEeqnarraybox*}[\IEEEeqnarraystrutmode \IEEEeqnarraystrutsizeadd{0.5pt} {0.5pt}][c]{,c/c,}
\mathbf{C}_\mathrm{1}&\mathbf{0}\\
\mathbf{0}&\mathbf{C}_\mathrm{2}%
\end{IEEEeqnarraybox*}\right),\quad
\mathbf{C}_\mathrm{SR} = \left(\begin{IEEEeqnarraybox*}[\IEEEeqnarraystrutmode \IEEEeqnarraystrutsizeadd{0.5pt} {0.5pt}][c]{,c/c,}
\mathbf{C}'_\mathrm{1}&\mathbf{0}\\
\mathbf{0}&\mathbf{C}'_\mathrm{2}%
\end{IEEEeqnarraybox*}\right)
\end{equation}
where $\mathbf{C}_\ell$ is a $M_\ell \times K_\ell$ matrix constructed at node $\mathrm{D}$ using the received coding vectors from node $\mathrm{S}_\ell$, and $\mathbf{C}'_\ell$ is a $M'_\ell \times K_\ell$ matrix that consists of the received coding vectors from node $\mathrm{S}_\ell$ at node $\mathrm{R}$. The dimensions of $\mathbf{C}_\mathrm{SD}$ and $\mathbf{C}_\mathrm{SR}$ are \mbox{$(M_1+M_2)\times(K_1+K_2)$} and \mbox{$(M'_1+M'_2)\times(K_1+K_2)$}, respectively.

In the second phase, if the relay node $\mathrm{R}$ successfully recovered the source packets of one or both source nodes, it linearly combines them in order to generate $N_\mathrm{R}$ coded packets. Therefore, the coding vector that accompanies each \mbox{relay-generated} coded packet consists of $K_1+K_2$ entries. If the relay node failed to decode the packets of either $\mathrm{S}_1$ or $\mathrm{S}_2$ then the first $K_1$ entries or the last $K_2$ entries of the coding vector, respectively, are set to zero. If $M_\mathrm{R}$ of the $N_\mathrm{R}$ transmitted coded packets are received by the destination node $\mathrm{D}$, a $M_\mathrm{R}\times (K_1+K_2)$ coding matrix $\mathbf{C}_\mathrm{RD}$ will be created and appended to $\mathbf{C}_\mathrm{SD}$. At the end of the second phase, the coding matrix at node $\mathrm{D}$ is
\begin{equation}
\mathbf{C}_\mathrm{D} = 
\left(\begin{IEEEeqnarraybox*}[\IEEEeqnarraystrutmode \IEEEeqnarraystrutsizeadd{0.5pt} {0.5pt}][c]{,c,}
\mathbf{C}_\mathrm{SD}\\ 
\mathbf{C}_\mathrm{RD}
\end{IEEEeqnarraybox*}\right) =
\left(\begin{IEEEeqnarraybox*}[\IEEEeqnarraystrutmode \IEEEeqnarraystrutsizeadd{0.5pt} {0.5pt}][c]{,c/c,}
\mathbf{C}_1&\mathbf{0}\\
\mathbf{0}&\mathbf{C}_2\\
\mathbf{C}_\mathrm{R1}&\mathbf{C}_\mathrm{R2}%
\end{IEEEeqnarraybox*}\right)
\end{equation}
which is a $(M_1+M_2+M_\mathrm{R})\times (K_1+K_2)$ block angular matrix. Note that $\mathbf{C}_\mathrm{RD}$ has been expressed as the concatenation of matrices $\mathbf{C}_\mathrm{R1}$ and $\mathbf{C}_\mathrm{R2}$, which were generated by node $\mathrm{R}$ and describe linear combinations of source packets originating from nodes $\mathrm{S}_1$ and $\mathrm{S}_2$, respectively. We stress that the coding vector transmitted by node $\mathrm{R}$ is twice the size of the coding vectors transmitted by $\mathrm{S}_1$ and $\mathrm{S}_2$; however, all coded packets in the network have the same size, which is customarily taken to be considerably larger than the size of the coding vectors.

The objective of this paper is to characterise the system performance of the considered two-source relay-aided network. More specifically, we will carry out a performance analysis to determine the probability that the destination node $\mathrm{D}$ will decode the $K_1+K_2$ source packets of both nodes $\mathrm{S}_1$ and $\mathrm{S}_2$, given that node $\mathrm{D}$ has recovered at least $K_1+K_2$ coded packets, that is, $(M_1+M_2+M_\mathrm{R})\geq K_1+K_2$. The impact of the chosen values for $N$ and $N_\mathrm{R}$ on the system performance will also be discussed.


\section{System Performance Analysis}
\label{sec:analysis}

Fundamental probabilities related to the rank of random matrices in $\mathrm{GF}(2)$ are summarised in this section and are subsequently used in the derivation of expressions for the probability that the destination node $\mathrm{D}$ will successfully recover the source packets of both source nodes, when they employ either non-systematic or systematic random linear network coding.

\subsection{Preliminaries: Fundamental Probability Expressions}

Let $\mathbf{X}$ be a $m\times k$ binary random matrix with $m\geq k$. We say that $\mathbf{X}$ is a \textit{full-rank} matrix if the rank of $\mathbf{X}$ is $k$ or, equivalently, $k$ of the $m$ rows of $\mathbf{X}$ are linearly independent. The probability of $\mathbf{X}$ being a full-rank matrix can be obtained as follows
\begin{equation}
\label{eq:prob_rect_full_rank}
\mathbb{P}(m,k) = \frac{f(m,k)}{2^{mk}}
\end{equation}
where $2^{mk}$ is the number of all $m\times k$ binary matrices and $f(m,k)$ is the number of all full-rank $m\times k$ binary matrices given by \cite{vanLint}
\begin{equation}
f(m,k) = \prod_{i=0}^{k-1}(2^{m}-2^{i}).
\end{equation}
The probability of $\mathbf{X}$ having rank $r\leq k$ when $m\geq r$ has also been derived in \cite{vanLint} and is equal to
\begin{equation}
\label{eq:prob_rect_rank_def}
\mathbb{P}_{r}(m,k) = 2^{-mk}\left(\frac{f(m,r)f(k,r)}{f(r,r)}\right).
\end{equation}
Note that for $r=k$, expression \eqref{eq:prob_rect_rank_def} reduces to \eqref{eq:prob_rect_full_rank}.

Let us now assume that matrix $\mathbf{X}$ has the following constrained structure 
\begin{equation}
\label{eq:def_block_angular}
\mathbf{X} = 
\left(\begin{IEEEeqnarraybox*}[\IEEEeqnarraystrutmode \IEEEeqnarraystrutsizeadd{0.5pt} {0.5pt}][c]{,c/c,}
\mathbf{A}&\mathbf{0}\\
\mathbf{0}&\mathbf{B}\\
\mathbf{C}&\mathbf{D}%
\end{IEEEeqnarraybox*}\right)
\end{equation}
where the dimensions of submatrices $\mathbf{A}$, $\mathbf{B}$, $\mathbf{C}$ and $\mathbf{D}$ are $a\times a'$, $b\times b'$, $c\times a'$ and $c\times b'$, respectively. Matrices of this type, which are known as \textit{block angular matrices}, were studied in \cite{Ferreira2013}. It was proven that the probability of $\mathbf{X}$ being full-rank is given by
\begin{equation}
\label{eq:prob_blockang_full_rank}
\mathbb{P}(a,a',b,b',c)\!=\!\!\!\!\!\!\!\!\!\!\sum_{i+j\geq a'+b'-c}\!\!\!\!\!\!\!\!\!\!\mathbb{P}_{i}(a,a')\mathbb{P}_{j}(b,b')\mathbb{P}(c,a'\!+\!b'\!-\!i\!-\!j).
\end{equation}As implied by \eqref{eq:prob_blockang_full_rank}, the rank of matrix $\mathbf{X}$ is $a'+b'$ if submatrix $\mathbf{A}$ has rank $i$, submatrix $\mathbf{B}$ has rank $j$ and the remaining \mbox{$a'+b'-i-j$} columns of $\mathbf{X}$ are linearly independent, for all valid values of $i$ and $j$.

Expressions \eqref{eq:prob_rect_full_rank}, \eqref{eq:prob_rect_rank_def} and \eqref{eq:prob_blockang_full_rank} will be invoked in the subsequent performance analysis. Note that character $\mathbb{P}$ is used exclusively to denote probabilities associated with the rank of matrices but character $P$ is used to refer to probabilities related to the system model under consideration.

\subsection{Decoding Probability for Non-systematic Network Coding}
\label{subsec:nonsys}

In the general case of point-to-point communication over a channel with erasure probability $p$, the probability of the receiving node recovering all of the $K$ source packets when $N$ coded packets have been transmitted can be written as follows
\begin{equation}
\label{eq:prob_PtP}
P(N,K,p)=\sum_{M=K}^{N}B(M,N,p)\;\mathbb{P}(M,K).
\end{equation}
The term $B(M,N,p)$ denotes the probability mass function of the binomial distribution, that is,
\begin{equation}
B(M,N,p)=\binom{N}{M}(1-p)^{M}p^{N-M}.
\end{equation}
Expression \eqref{eq:prob_PtP} enumerates all possible scenarios of retrieving the $K$ source packets when $M\geq K$ coded packets have been successfully received and have formed a full-rank $M\times K$ coding matrix. 

In the particular case of the considered relay-aided network, the probability that the destination node $\mathrm{D}$ will recover the source packets of both source nodes can be decomposed into the following three components:

\subsubsection{Unaided communication}

Even though the relay node $\mathrm{R}$ has been deployed in the network, the destination node $\mathrm{D}$ could recover all of the source packets without the help of node $\mathrm{R}$. The implies that both submatrices $\mathbf{C}_1$ and $\mathbf{C}_2$ in \eqref{eq:codmtx_SD_SR} are full-rank matrices and, consequently, $\mathbf{C}_\mathrm{SD}$ is also a full-rank matrix. Therefore, the probability that node $\mathrm{D}$ will recover the $K_1+K_2$ source packets based solely on the $N_1+N_2$ transmitted coded packets can be obtained using \eqref{eq:prob_PtP} as follows
\begin{equation}
\label{eq:D_without_R}
P_{\mathrm{S}}=P(N_1,K_1,p_{1,\mathrm{D}})\;P(N_2,K_2,p_{2,\mathrm{D}}).
\end{equation}

\subsubsection{Partially aided communication}

In this mode, the destination node recovers the $K_\ell$ source packets of node $\mathrm{S}_\ell$ based on coded packets transmitted both via the relay node and over the direct link between $\mathrm{S}_\ell$ and $\mathrm{D}$. The destination node retrieves the source packets of the other source node, denoted by $\mathrm{S}_{\bar{\ell}}$ where $\bar{\ell}=1,2$ and $\bar{\ell}\neq\ell$, without the assistance of the relay node. The probability that node $\mathrm{D}$ will recover the $K_1+K_2$ source packets, when transmission from node $\mathrm{S}_\ell$ is aided by the relay node $\mathrm{R}$ while transmission from node $\mathrm{S}_{\bar{\ell}}$ is unaided, can be upper-bound by the following product
\begin{equation}
\label{eq:D_with_R_and_Sj}
\begin{split}
P_{\mathrm{S}_{\ell}\mathrm{R}\mathrm{D}}\leq\;&P(N_{\bar{\ell}},K_{\bar{\ell}},p_{\bar{\ell},\mathrm{D}})\:P(N_{\ell},K_{\ell},p_{\ell,\mathrm{R}})\\
&\cdot\sum_{M_{\ell}=0}^{N_{\ell}}B(M_{\ell},N_{\ell},p_{\ell,\mathrm{D}})\\
&\cdot\!\!\!\!\!\!\!\!\!\sum_{i=0}^{\min(M_\ell,K_{\ell}-1)}\!\!\!\!\!\!\!\mathbb{P}_i(M_\ell,K_{\ell})P(N_\mathrm{R}, K_{\ell}-i, p_{\mathrm{R,D}}).
\end{split}
\end{equation}
The first two terms on the right-hand side of \eqref{eq:D_with_R_and_Sj} represent the probability that nodes $\mathrm{D}$ and $\mathrm{R}$ will recover the source packets of nodes $\mathrm{S}_{\bar{\ell}}$ and $\mathrm{S}_{\ell}$, respectively, when the direct links are used. The remaining two lines compute the probability that node $\mathrm{D}$ will construct a coding matrix of rank $K_\ell$ by obtaining $i$ linearly independent coding vectors from node $\mathrm{S}_{\ell}$ and $K_{\ell}-i$ linearly independent coding vectors from node $\mathrm{R}$. Derivation of this probability invoked and extended a degraded version of the right-hand side of \eqref{eq:prob_blockang_full_rank}, where $\mathbf{X}$ in \eqref{eq:def_block_angular} was redefined as $\mathbf{X}=\left(\mathbf{A}\;\;\mathbf{C}\right)^\intercal$.

The reason that the right-hand side of \eqref{eq:D_with_R_and_Sj} is an upper bound and not the exact expression for $P_{\mathrm{S}_{\ell}\mathrm{R}\mathrm{D}}$ lies to the fact that the probability of the relay node decoding the packets of node $\mathrm{S}_{\ell}$ is \textit{not independent} of the probability that the destination node will decode the packets of the same node. For example, consider the case when $N_{\ell}=10$ coded packets are transmitted to both $\mathrm{D}$ and $\mathrm{R}$ and $p_{\ell,\mathrm{D}}=p_{\ell,\mathrm{R}}=0.1$. If each node recovers $9$ coded packets then each node will have at least $8$ of them in common. Therefore, if node $\mathrm{D}$ fails to recover the source packets of node $\mathrm{S}_{\ell}$, node $\mathrm{R}$ will most likely also fail to recover them and will not be in the position to assist node $\mathrm{S}_{\ell}$ in its transmission. However, as the value of the product $N_{\ell}\:p_{\ell,\mathrm{R}}$ or $N_{\ell}\:p_{\ell,\mathrm{D}}$ increases, the upper bound gets tighter, as will become evident in Section \ref{sec:results}.

Using \eqref{eq:D_with_R_and_Sj}, the probability that the destination node will recover the source packets of both $\mathrm{S}_1$ and $\mathrm{S}_2$, when either $\mathrm{S}_1$ or $\mathrm{S}_2$ is aided by the relay node $\mathrm{R}$, is given by
\begin{equation}
\label{eq:D_with_SR}
P_{\mathrm{SRD}} = P_{\mathrm{S}_{1}\mathrm{R}\mathrm{D}} + P_{\mathrm{S}_{2}\mathrm{R}\mathrm{D}}.
\end{equation}

\subsubsection{Fully aided communication}

In this case, both $\mathrm{S}_1$ and $\mathrm{S}_2$ need the aid of the relay node $\mathrm{R}$ in order to deliver the necessary number of coded packets to the destination node. Node $\mathrm{D}$ successfully decodes the coded packets transmitted via node $\mathrm{R}$ and over the two direct links, and recovers all source packets. The probability that node $\mathrm{D}$ will recover the $K_1+K_2$ source packets, when both source nodes are assisted by the relay node, can be upper-bound as follows
\begin{equation}
\label{eq:D_with_R}
\begin{split}
P_{\mathrm{RD}}&\leq P(N_1,K_1,p_{1,\mathrm{R}})\: P(N_2,K_2,p_{2,\mathrm{R}})\\
&\!\!\cdot\sum_{M_{1}=0}^{N_1}B(M_1,N_1,p_{1,\mathrm{D}})\sum_{M_{2}=0}^{N_2}B(M_2,N_2,p_{2,\mathrm{D}})\\
&\!\!\cdot\sum_{i=0}^{i_{\max}}\:\sum_{j=0}^{j_{\max}}\mathbb{P}_i(M_1,K_1)\mathbb{P}_j(M_2,K_2)\\
&\!\!\cdot\:P(N_\mathrm{R},\:K_1\!+\!K_2\!-\!i\!-\!j,\:p_{\mathrm{R,D}}).
\end{split}
\end{equation}
The first line on the right-hand side of \eqref{eq:D_with_R} expresses the probability that node $\mathrm{R}$ will recover the source packets of both $\mathrm{S}_1$ and $\mathrm{S}_2$. The remaining lines compute the probability that node $\mathrm{D}$ will receive $i$, $j$ and $K_1+K_2-i-j$ linearly independent coding vectors from $\mathrm{S}_1$, $\mathrm{S}_2$ and $\mathrm{R}$, respectively, for all valid values of $i$ and $j$. Similarly to \eqref{eq:D_with_R_and_Sj}, we set the upper limit of the third sum in \eqref{eq:D_with_R} equal to \mbox{$i_{\max}=\min(M_1, K_{1}-1)$}; this ensures that the number of linearly independent coded vectors $i$, which have been received directly from node $\mathrm{S}_1$, is neither greater than the total number of received coded vectors $M_1$, nor equal to or greater than the number of source packets $K_1$. The definition of $i_{\max}$ prevents $i$ from taking the value $K_1$ because cases where node $\mathrm{D}$ can recover the $K_1$ source packets without the help of node $\mathrm{R}$ have already been considered in unaided and partially aided communication. Following a similar line of reasoning, we set $j_{\max}=\min(M_2, K_{2}-1)$ in \eqref{eq:D_with_R}. Observe that the last two lines of \eqref{eq:D_with_R} constitute a formula that is a constrained extension of \eqref{eq:prob_blockang_full_rank}.

The overall decoding probability at the destination node $\mathrm{D}$ can be obtained by adding the three constituent probabilities, that is,
\begin{equation}
\label{eq:dec_prob}
P_{\mathrm{D}} = P_{\mathrm{S}} + P_{\mathrm{SRD}} + P_{\mathrm{RD}}.
\end{equation}
We remark that if the right-hand side of \eqref{eq:D_without_R}, \eqref{eq:D_with_R_and_Sj} and \eqref{eq:D_with_R} are used in \eqref{eq:dec_prob} to compute $P_{\mathrm{S}}$, $P_{\mathrm{SRD}}$ and $P_{\mathrm{RD}}$, respectively, an upper bound on $P_{\mathrm{D}}$ will be obtained. 

\subsection{Decoding Probability for Systematic Network Coding}

In \cite{Jones2015}, systematic network coding for point-to-point communication was studied and it was proven that the probability of a receiving node decoding all of the $K$ source packets, given that $K\leq M \leq N$ packets have been successfully received, is
\begin{equation}
\label{eq:sys_prob_rect_full_rank}
P'(M,K,N)=\frac{\sum_{h=h_{\min}}^{K}\!\binom{K}{h}\binom{N-K}{M-h}\:\mathbb{P}(M\!-\!h,K\!-\!h)}{\binom{N}{M}}
\end{equation}
where $h_{\min}=\max(0,M-N+K)$. Expression \eqref{eq:sys_prob_rect_full_rank} considers the possibility of receiving $h$ systematic and, hence, linearly independent packets out of the $K$ transmitted systematic packets and computes the probability that there exist $K-h$ linearly independent coded packets among the remaining $M-h$ packets, for all valid values of $h$. Following the same line of reasoning as in \cite{Jones2015}, we can express the probability of receiving $r\leq K$ linearly independent coded packets as
\begin{equation}
\label{eq:sys_prob_rect_rank_def}
P'_{\!r}(M,\!K,\!N)\!=\!\frac{\sum_{h=h_{\min}}^{r}\!\!\binom{\!K\!}{\!h\!}\binom{N\!-\!K}{M\!-\!h}\mathbb{P}_{r-h}(M\!-\!h,K\!-\!h)}{\binom{N}{M}}
\end{equation}provided that $M\geq r$. Similarly to the case of non-systematic network coding, the probability of the receiving node recovering all of the $K$ source packets when $N$ packets have been transmitted, denoted by $P'(N,K,p)$, can be obtained from \eqref{eq:prob_PtP} by replacing $\mathbb{P}(M,K)$ with $P'(M,K,N)$.

Taking into account \eqref{eq:sys_prob_rect_full_rank} and \eqref{eq:sys_prob_rect_rank_def} and using the same train of thought as in Section \ref{subsec:nonsys}, we can obtain an expression for the performance of the considered two-source single-relay network for the case of systematic network coding. More specifically, the probability that the destination node will recover the source packets of both source nodes is given by 
\begin{equation}
\label{eq:sys_dec_prob}
P'_{\mathrm{D}} = P'_{\mathrm{S}} + \left(P'_{\mathrm{S_{1}RD}}+P'_\mathrm{S_{2}RD}\right) + P'_{\mathrm{RD}}
\end{equation}
where
\begin{equation}
\label{eq:sys_D_without_R}
\hspace{-12mm}P'_{\mathrm{S}}=P'(N_1,K_1,p_{1,\mathrm{D}})\;P'(N_2,K_2,p_{2,\mathrm{D}}),
\end{equation}
\begin{equation}
\label{eq:sys_D_with_R_and_Sj}
\begin{split}
P'_{\mathrm{S}_{\ell}\mathrm{R}\mathrm{D}}&\leq\;P'(N_{\bar{\ell}},K_{\bar{\ell}},p_{\bar{\ell},\mathrm{D}})\;P'(N_{\ell},K_{\ell},p_{\ell,\mathrm{R}})\\
&\cdot\sum_{M_{\ell}=0}^{N_{\ell}}B(M_{\ell},N_{\ell},p_{\ell,\mathrm{D}})\\
&\cdot\!\!\!\!\!\!\!\!\!\sum_{i=0}^{\min(M_\ell,K_{\ell}-1)}\!\!\!\!\!\!\!P'_i(M_\ell,K_{\ell},N_{\ell})\;P(N_{\mathrm{R}},K_{\ell}-i
,p_{\mathrm{R,D}})
\end{split}
\end{equation}for $\ell=1,2$, and
\begin{equation}
\label{eq:sys_D_with_R}
\begin{split}
P'_{\mathrm{RD}}\leq\;&P'(N_1,K_1,p_{1,\mathrm{R}})\:P'(N_2,K_2,p_{2,\mathrm{R}})\\
&\!\!\cdot\sum_{M_{1}=0}^{N_1}B(M_1,N_1,p_{1,\mathrm{D}})\sum_{M_{2}=0}^{N_2}B(M_2,N_2,p_{2,\mathrm{D}})\\
&\!\!\cdot\sum_{i=0}^{i_{\max}}\:\sum_{j=0}^{j_{\max}}P'_i(M_1,K_1,N_1)P'_j(M_2,K_2,N_2)\\
&\!\!\cdot\:P(N_{\mathrm{R}},\:K_1\!+\!K_2\!-\!i\!-\!j,\:p_{\mathrm{R,D}}).
\end{split}
\end{equation}

The validity and tightness of the derived performance bounds will be investigated in the following section.

\section{Results and Discussion}
\label{sec:results}

In this section, comparisons between the derived theoretical upper bounds and simulation results will be carried out for both systematic and non-systematic network coding. For convenience, a symmetric network configuration has been considered, according to which $K_1=K_2=K$, $N_1=N_2=N$, $p_{1,\mathrm{D}}=p_{2,\mathrm{D}}=p_{\mathrm{S},\mathrm{D}}$ and $p_{1,\mathrm{R}}=p_{2,\mathrm{R}}=p_{\mathrm{S},\mathrm{R}}$.

\begin{figure}[t]
\centering
\includegraphics[width=0.95\columnwidth]{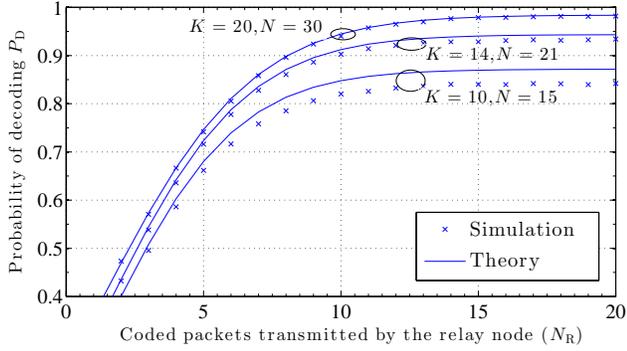}
\caption{Comparison between theoretical upper bounds obtained from \eqref{eq:dec_prob} and simulation results for different values of $K$ and $N$. The erasure probabilities have been set to $p_{\mathrm{S},\mathrm{D}}=0.3$, $p_{\mathrm{S},\mathrm{R}}=0.1$ and $p_{\mathrm{R},\mathrm{D}}=0.2$.}
\vspace{-4mm}
\label{fig:fig_dependency}
\end{figure}

Fig.~\ref{fig:fig_dependency} compares simulation results with the theoretical expression in \eqref{eq:dec_prob} as a function of $N_\mathrm{R}$, for different values of $K$ and $N$. As explained in Section \ref{subsec:nonsys}, the interdependency between the decoding probability at node $\mathrm{R}$ and the decoding probability at node $\mathrm{D}$ is evident when $K=10$ and $N=15$; in this case, the upper bound yields a marginally higher decoding probability than that obtained via simulations. However, the interdependency becomes smaller and the upper bound gets tighter with an increasing number of source packets $K$ and, consequently, an increasing number of transmitted packets $N$. We observe that for $K=20$ and $N=30$, the derived upper bound coincides with the simulation results.

\begin{figure}[t]
\centering
\includegraphics[width=0.95\columnwidth]{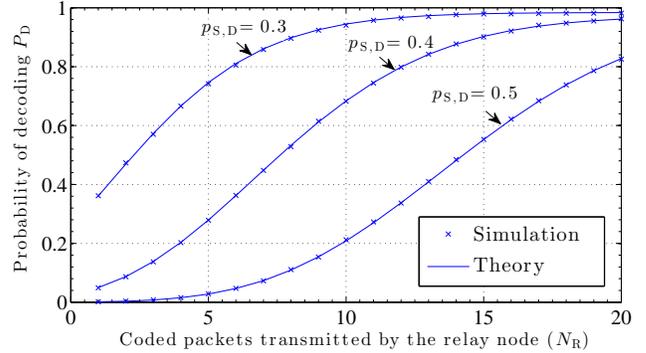}
\vspace{-0.8mm}
\caption{Comparison between theoretical upper bounds obtained from \eqref{eq:dec_prob} and simulation results for different values of $p_{\mathrm{S},\mathrm{D}}$. The remaining system parameters have been set to $K=20$, $N=30$, $p_{\mathrm{S},\mathrm{R}}=0.3$ and $p_{\mathrm{R},\mathrm{D}}=0.2$.}
\vspace{-2mm}
\label{fig:decoding_prob}
\end{figure}

The tightness of the proposed upper bound is also illustrated in Fig.~\ref{fig:decoding_prob}, which depicts the impact of the source-to-destination channel quality, represented by $p_{\mathrm{S,D}}$, and the number of coded packets $N_\mathrm{R}$ transmitted by the relay node on the system decoding probability $P_\mathrm{D}$. As expected, aid by the relay is of key importance to the source nodes as the quality of the direct channel between each source node and the destination node deteriorates. The theoretical bounds accurately quantify the relationship between $p_{\mathrm{S,D}}$ and the number of coded packets $N_{\mathrm{R}}$ that need to be transmitted by the relay to achieve a target decoding probability.

\begin{figure}[t]
\centering
\includegraphics[width=0.93\columnwidth]{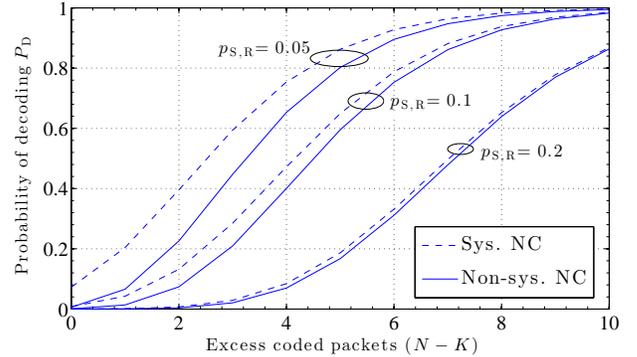}
\caption{Performance comparison of systematic and non-systematic network coding as a function of the excess coded packets $N-K$ transmitted by each source node for various values of $p_{\mathrm{S},\mathrm{R}}$. The remaining system parameters have been set to $K=20$, $N_\mathrm{R}=15$, $p_{\mathrm{S},\mathrm{D}}=0.3$ and $p_{\mathrm{R},\mathrm{D}}=0.1$.}
\label{fig:sys_decoding}
\vspace{-4mm}
\end{figure}

Fig.~\ref{fig:sys_decoding} carries out a performance comparison between systematic and non-systematic network coding (NC) for various values of $p_{\mathrm{S,R}}$. As is evident from the figure, if systematic NC is used at the source nodes and the source-to-relay channel conditions are good, the destination node requires fewer excess coded packets $N-K$ from the source nodes to correctly decode all of the $K_1+K_2$ source packets. This observation is in agreement with the findings in \cite{Jones2015} for point-to-point communication. As the source-to-relay channel quality deteriorates, systematic NC performs similarly to non-systematic NC. Nevertheless, systematic NC still offers the benefits of progressive packet recovery and reduced decoding complexity, as detailed in \cite{Jones2015}.

\section{Conclusions and Future Work}
\label{sec:concl}

This paper studied the performance of a network comprising two source nodes transmitting to a destination node via a relay node, where random linear network coding is used both at the source nodes and the relay node. Upper bounds on the probability of the destination node successfully recovering the packets of both source nodes were derived for both systematic and non-systematic network coding. Simulation results confirmed the validity of our theoretical analysis and established that the upper bounds get tighter and accurately predict the system decoding probability for an increasing number of transmitted coded packets by the source nodes. Furthermore, we demonstrated that systematic network coding can yield a similar or better performance than non-systematic network coding depending on the quality of the uplink channels.

Future work will extend the system model to networks of multiple source nodes and, possibly, multiple relay nodes, and will aim to obtain expressions for the overall decoding probability. In this endeavour, both binary and non-binary network coding will be considered, that is, coded symbols in transmitted packets will be elements of a finite field $\mathrm{GF}(q)$ of size $q\geq 2$. An additional line of investigation to be pursued in the short term is the optimisation of the ratio between the coded packets transmitted by each source node and the coded packets transmitted by the relay node in order to achieve a target decoding probability, while meeting the energy constraints of the source nodes. We will also strive to draw comparisons of the considered decode-and-forward scheme with other network-coded forwarding protocols as well as with networks employing physical network coding.

\section*{Acknowledgment}
This work was conducted as part of the R2D2 project, which is supported by the Engineering and Physical Sciences Research Council (EPSRC) under Grant EP/L006251/1. The first author would also like to acknowledge Lancaster University for offering him a fully-funded PhD studentship.

\IEEEtriggeratref{8}
\bibliographystyle{IEEEtran}
\bibliography{ICC15WSPbib}

\begin{thebibliography}{10}
\providecommand{\url}[1]{#1}
\csname url@samestyle\endcsname
\providecommand{\newblock}{\relax}
\providecommand{\bibinfo}[2]{#2}
\providecommand{\BIBentrySTDinterwordspacing}{\spaceskip=0pt\relax}
\providecommand{\BIBentryALTinterwordstretchfactor}{4}
\providecommand{\BIBentryALTinterwordspacing}{\spaceskip=\fontdimen2\font plus
\BIBentryALTinterwordstretchfactor\fontdimen3\font minus
  \fontdimen4\font\relax}
\providecommand{\BIBforeignlanguage}[2]{{%
\expandafter\ifx\csname l@#1\endcsname\relax
\typeout{** WARNING: IEEEtran.bst: No hyphenation pattern has been}%
\typeout{** loaded for the language `#1'. Using the pattern for}%
\typeout{** the default language instead.}%
\else
\language=\csname l@#1\endcsname
\fi
#2}}
\providecommand{\BIBdecl}{\relax}
\BIBdecl

\bibitem{Ahlswede2000}
R.~Ahlswede, N.~Cai, S.~R. Li, and R.~W. Weung, ``Network information flow,''
  \emph{{IEEE} Trans. on Inform. Theory}, vol.~46, no.~4, pp. 1204--1216, Jul.
  2000.

\bibitem{Seferoglu2011}
H.~Seferoglu, A.~Markopoulou, and K.~K. Ramakrishnan,
  ``{I\textsuperscript{2}NC}: Intra- and inter-session network coding for
  unicast flows in wireless networks,'' in \emph{Proc. IEEE Conf. on Computer
  Commun. (INFOCOM)}, Shanghai, China, Apr. 2011.

\bibitem{Bao2005}
X.~Bao and J.~Li, ``Matching code-on-graph with network-on-graph: Adaptive
  network coding for wireless relay networks,'' in \emph{Proc. Allerton Conf.
  on Commun., Control and Computing}, Monticello, USA, Sep. 2005.

\bibitem{Chen2006}
Y.~Chen, S.~Kishore, and J.~Li, ``Wireless diversity through network coding,''
  in \emph{Proc. IEEE Wireless Commun. and Networking Conf. (WCNC)}, Las Vegas,
  USA, Apr. 2006.

\bibitem{Woldegebreal2007}
D.~H. Woldegebreal and H.~Karl, ``Multiple-access relay channel with network
  coding and non-ideal source-relay channels,'' in \emph{Proc. Int. Symp. on
  Wireless Commun. Systems (ISWCS)}, Trondheim, Norway, Oct. 2007.

\bibitem{Hausl2006}
C.~Hausl and P.~Dupraz, ``Joint network-channel coding for the multiple-access
  relay channel,'' in \emph{Proc. IEEE Int. Conf. on Sensor, Mesh and Ad Hoc
  Commun. and Networks (SECON)}, Reston, USA, Sep. 2006.

\bibitem{Yang2007}
S.~Yang and R.~Koetter, ``Network coding over a noisy relay: a belief
  propagation approach,'' in \emph{Proc. IEEE Int. Symp. on Inform. Theory
  (ISIT)}, Nice, France, Jun. 2007.

\bibitem{Nokleby2012}
M.~Nokleby, B.~Nazer, B.~Aazhang, and N.~Devroye, ``Relays that cooperate to
  compute,'' in \emph{Proc. Int. Symp. on Wireless Commun. Systems (ISWCS)},
  Paris, France, Aug. 2012.

\bibitem{Molisch2007}
A.~F. Molisch, N.~B. Mehta, J.~S. Yedidia, and J.~Zhang, ``Performance of
  fountain codes in collaborative relay networks,'' \emph{{IEEE} Trans. on
  Wireless Commun.}, vol.~11, no.~6, pp. 4108--4119, Nov. 2007.

\bibitem{Tarable2009}
A.~Tarable and I.~Chatzigeorgiou, ``Randomly select and forward: {E}rasure
  probability analysis of a probabilistic relay channel model,'' in \emph{Proc.
  IEEE Inform. Theory Workshop (ITW)}, Taormina, Italy, Oct. 2009.

\bibitem{Kurniawan2010}
E.~Kurniawan, S.~Sumei, K.~Yen, and K.~Chong, ``Network coded transmission of
  fountain codes over cooperative relay networks,'' in \emph{Proc. IEEE
  Wireless Commun. and Networking Conf. (WCNC)}, Sydney, Australia, Apr. 2010.

\bibitem{Ferreira2013}
P.~J. S.~G. Ferreira, B.~Jesus, J.~Vieira, and A.~J. Pinho, ``The rank of
  random binary matrices and distributed storage applications,'' \emph{{IEEE}
  Commun. Lett.}, vol.~17, no.~1, pp. 151--154, Jan. 2013.

\bibitem{Fitzek2013}
J.~Krigslund, J.~Hansen, M.~Hundeboll, D.~Lucani, and F.~Fitzek, ``{CORE}:
  {COPE} with {MORE} in wireless meshed networks,'' in \emph{Proc. IEEE 77th
  Vehicular Technology Conference (VTC Spring)}, Jun. 2013.

\bibitem{vanLint}
J.~H. van Lint and R.~M. Wilson, \emph{A Course in Combinatorics},
  2nd~ed.\hskip 1em plus 0.5em minus 0.4em\relax Cambridge University Press,
  2001.

\bibitem{Jones2015}
A.~L. Jones, I.~Chatzigeorgiou, and A.~Tassi, ``Binary systematic network
  coding for progressive packet decoding,'' in \emph{Proc. IEEE Intern. Conf.
  on Commun. (ICC)}, London, UK (to appear), Jun. 2015.

\end{thebibliography}

\end{document}